\documentclass[epsf,12pt]{article}

\usepackage{amsmath,amssymb,theorem}

\newcommand{\R}{{\mathbb{R}}}
\newcommand{\Z}{{\mathbb{Z}}}
\newcommand{\C}{{\mathbb{C}}}

\newcommand{\ba}{\begin{array}}
\newcommand{\ea}{\end{array}}

\newcommand{\bp}{\begin{pmatrix}}
\newcommand{\ep}{\end{pmatrix}}

\newcommand{\bps}{\begin{smallmatrix}}
\newcommand{\eps}{\end{smallmatrix}}

\newcommand{\bi}{\begin{itemize}}
\newcommand{\ei}{\end{itemize}}

\newcommand{\cD}{{\cal D}}

\newcommand{\cP}{{\cal P}}

\newcommand{\x}{\xi}

\newcommand{\del}{\partial}

\newcommand{\fr}{\frac}
\newcommand{\half}{\frac{1}{2}}
\newcommand{\qua}{\frac{1}{4}}

\newcommand{\Dsl}{\mbox{\ooalign{\hfil/\hfil\crcr$D$}}}

\newcommand{\mn}{{\mu\nu}}
\newcommand{\nn}{\nonumber\\}

\newcommand{\Ah}{\hat{A}}
\newcommand{\Ch}{\hat{C}}
\newcommand{\Dh}{\hat{D}}    
\newcommand{\Fh}{\hat{F}}

\newcommand{\Th}{\hat{T}}

\newcommand{\delh}{\hat{\partial}}
\newcommand{\Omegah}{\hat{\Omega}}

\def \cD{{\cal D}}

\def \cDb{{\bar {\cal D}}}

\DeclareMathOperator{\Tr}{Tr}

\def \Re{\mathrm{Re}}
\def \Im{\mathrm{Im}}

\def \rank{\mathrm{rank}}

\def \ch{\mbox{ch}}

\def \half{\frac{1}{2}}

\def \Eh{\hat{E}}

\def \l{{\frak l}}

\def \l({\left(}
\def \r){\right)}

\def \0{{\bf 0}}
\def \1{{\bf 1}}

{\theorembodyfont{\rmfamily}

}

\begin{document}

\begin{titlepage}
\thispagestyle{empty}
\begin{flushright}
hep-th/0208059\\
UT-02-40\\
August, 2002 \\
\end{flushright}

\vskip 1.5 cm

\begin{center}
\noindent{\LARGE Gauge Fields on Tori and T-duality}\\
\noindent{
 }\\
\renewcommand{\thefootnote}{\fnsymbol{footnote}}

\vskip 2cm

{\large Masashi Hamanaka\footnote{e-mail address:
    hamanaka@hep-th.phys.s.u-tokyo.ac.jp} and 
    Hiroshige Kajiura\footnote{e-mail address:
    kuzzy@ms.u-tokyo.ac.jp}}
\vspace{15mm}

${}^{*}${\it Department of Physics, University of Tokyo,\\ 
Hongo 7-3-1, Bunkyo-ku, Tokyo 113-0033, Japan}

\vspace{6mm}

${}^{\dag}${\it Graduate School of Mathematical Sciences, 
University of Tokyo,\\
Komaba 3-8-1, Meguro-ku, Tokyo 153-8914, Japan}

\vskip 1.5cm
\end{center}
\begin{abstract}
We discuss gauge fields on tori in diverse dimensions, 
mainly in two and four dimensions.
We construct various explicit gauge fields which have
some topological charges
and find the Dirac zero modes in the background
of the gauge fields.
By using the zero mode,
we give new gauge fields on the dual torus,
which is a gauge theoretical description of T-duality 
transformation of the corresponding D-brane systems
including D\=D systems. 
{}From the transformation,
we can easily see the duality 
expected from the index theorem. 
It is also mentioned that, for each topological charges, the 
corresponding constant curvature bundle can be constructed 
and their duality transformation can be performed 
in terms of Heisenberg modules. 

\end{abstract}
\vfill
 
\end{titlepage}
\vfill
\setcounter{footnote}{0}
\renewcommand{\thefootnote}{\arabic{footnote}}
\newpage

\baselineskip 6mm

\section{Introduction}

In the study of non-perturbative aspects of field theories,
solitons and instantons have played crucial roles.
Especially, gauge theories on tori possess various
topological characters which comes from the non-triviality
of the topology of the tori, 
and have been studied intensively,
for example,
in order to explain quark confinement \cite{tHooft, Gonzalez}.

Gauge theories on four-dimensional tori $T^4$ have the mysterious duality,
that is, {\it Fourier-Mukai-Nahm duality},
which is one-to-one correspondence 
(more strongly hyperK\"ahler isometry) 
between instanton moduli space on the torus $T^4$ and
that on the dual torus $\Th^4$ \cite{Mukai, Schenk, BvB, DoKr}.
The duality transformation is called {\it Nahm transformation}
which originated in the application of ADHM construction \cite{ADHM}
to BPS monopoles \cite{Nahm}.
One of the interesting points is that
under the Nahm transformation, 
the rank of the gauge group and
the instanton number 
are interchanged.
Physically, the transformation is just the T-duality transformation. 

Gauge theoretical analysis are often strong to
study D-brane dynamics, such as tachyon condensation \cite{Harvey}.
Nahm transformation is taken for the gauge fields on D-brane
and hence is expected to describe more in detail in some points
than usual T-duality transformation
and to reveal new aspects of D-brane dynamics.

In the present paper, we study gauge theories on tori
in diverse dimensions and take explicit Nahm-like transformations
via the Dirac zero modes in the background of the gauge fields.
We construct various explicit gauge fields which have
constant curvatures.
This specifies the corresponding D-brane systems. 
We find the explicit Dirac zero modes in the background
of the gauge fields which show consistency
with the family index theorem \cite{AtSi}.
By using the zero mode,
we give new gauge fields on the dual torus in canonical ways,
which is consistent with the T-duality
transformation of the corresponding D-brane systems
including D\=D systems.
We focus on two- and four-dimensional tori.
The higher-dimensional extension is straightforward. 

This paper is organized as follows.
In section 2, we discuss gauge theories on two-dimensional tori and 
T-duality transformation of D0-D2 brane systems.
In section 3, we discuss gauge theories on four-dimensional tori and 
T-duality transformation of D0-D4 brane systems, which is Nahm transformation.
In section 4, we discuss the higher-dimensional extensions.
Finally in section 5, we give conclusions and 
some comments on noncommutative extensions of the present discussion.

\section{Gauge fields on $T^2$ and T-duality}

In this section, we discuss gauge fields on two-tori $T^2$
and the gauge theoretical descriptions of T-duality transformations
of D0-D2 brane systems.
We first give a general scheme of two-dimensional version of 
Nahm transformation and then apply it to
some explicit gauge fields.
The key point in the transformation is to find the Dirac zero mode.
The result shows a beautiful duality and is consistent with T-duality.
We mainly treat a flux solution where the rank of the gauge field
is $N$ and the first Chern number\footnote{In the present paper, we use
the word ``the $n$-th Chern number'' as the integral 
of the $n$-th Chern character.} (the magnetic flux) is $k$
which corresponds to $k$ D0-branes and $N$ D2-branes on $T^2$. 
Finally we comment on the generalization to non-flux solutions 
including  D0-\=D0 branes on D2-branes on $T^2$.

For simplicity, we set the both periods of the torus $T^2$ $2\pi$.

\vspace{2mm}

First of all, we shall recall general results of Atiyah-Singer 
family index theorem.
The detailed discussion is given later soon.

Main theorem is:
\begin{eqnarray}
\label{f_index}
\ch({\mbox{ind}} \Dsl_\xi)=\int_{T^{2n}}\ch(\cP)\ch(E),
\end{eqnarray}
where $\ch(E):=\Tr \exp(F/2\pi i)$ 
represents the Chern character of the vector bundle $E$
over a $2n$-dimensional torus and $F$ is the curvature two form of $E$.
The $i$-th Chern character $\ch_i(E)$ is concretely represented as
\begin{eqnarray*}
\ch_1(E)=\fr{i}{2\pi}\Tr F,~~~\ch_2(E)=-\fr{1}{8\pi^2}\Tr F\wedge F,\cdots.
\end{eqnarray*}
$\cP$ are called 
the Poincar\'e bundle.
On even dimensional tori, the Dirac operator $\Dsl_\xi$
is decomposed into two Weyl components $\cD_\xi$ and $\cDb_\xi$. 
$\mbox{ind} \Dsl_\xi$ is then defined as 
$\mbox{ind} \Dsl_\xi:=\ker\cDb_\xi-\ker \cD_\xi$, which belong to 
$K_0$-group over the dual torus $\Th^{2n}$. 
If the dimension of either $\ker\cDb_\xi$ or $\ker \cD_\xi$ is constant 
with respect to $\xi$, the index theorem implies the dimensions 
of both $\ker\cDb_\xi$ and $\ker \cD_\xi$ are constant, and then 
$\ker\cDb_\xi$ and $\ker \cD_\xi$ are vector bundles over dual torus. 
Physically this is just the D\=D system. In particular 
if $\ker \cD_\xi$ is trivial, the index bundle $\mbox{ind}\Dsl_\xi$ becomes
so called zero-mode bundle $\ker\cDb_\xi=:\Eh$ 
over the dual torus. 
\vspace{2mm}

Now we define the two-dimensional Nahm transformation explicitly.
Suppose that
the gauge group is $U(N)$ and 
the first Chern number is $k$.
The coordinates of torus $T^2$ and the dual torus 
are denoted by $(x_1,x_2)$ and $(\xi_1,\xi_2)$, respectively.
The indices run as follows: 
space indices $\mu,\nu=1,2$;
D2 indices $u,v=1,\ldots,N$;
D0 indices $p,q=1,\ldots,k$.

Let us consider two-dimensional (massless) Dirac operator on $T^2$:
\begin{eqnarray*}
\Dsl:=\gamma_\mu \otimes D_{\mu}=\gamma_\mu \otimes (\del_\mu+A_\mu),
\end{eqnarray*}
where $\gamma_\mu=-i\sigma_\mu, ~\mu=1,2$. 
($\sigma_i$ are usual Pauli matrices.)
The Dirac operator act on the section of the tensor product of
the vector bundle $E$ and spinor bundle $S$ on $T^2$.

Now we take the tensor product of (the pull-back of) $E$ 
and the Poincar\'e bundle $P$ on $T^2 \times \Th^2$
The connection of the Poincar\'e bundle is given by $-i\xi_\mu$.
Hence the connection of this tensor bundle 
is $D_{\mu}(\xi)=\del_\mu+A_\mu-i\xi_\mu$
and the Dirac operator is as follows:
\begin{eqnarray*}
\Dsl_\xi &:=& \gamma_\mu \otimes D_{\mu}(\xi)=
\left(
\ba{cc}
0&\cD_\xi\\\cDb_\xi&0
\ea
\right)\\ 
\Dsl_\xi^2 &=&
\left(
\ba{cc}
\cD_\xi\cDb_\xi&0\\0&\cDb_\xi\cD_\xi
\ea
\right).
\end{eqnarray*}
Let us suppose that the (chiral-decomposed)
Dirac operator $\cD_\xi$ has no smooth solution.
Then the (family) index theorem says that 
the Dirac operator $\cDb_\xi$ has independent
normalized $k$ solutions:
\begin{eqnarray}
\label{dirac_02}
\cDb_\xi \psi =0,
\end{eqnarray}
where $\psi(x,\xi)$ is $N\times k$ matrix 
in which the each row vector $\psi^p$
is the independent solution and is called the Dirac zero mode.
The gauge field on the dual torus can be constructed by
the orthonormal projection onto the zero-mode bundle:
$
\Dh_\mu= P \delh_\mu=(\psi\psi^\dagger)\delh_\mu
$, or explicitly,
\begin{eqnarray}
\label{gauge_02}
\Ah_\mu(\xi) = \int_{T^2}d^2x~ \psi^\dagger\delh_\mu\psi. 
\end{eqnarray}
The dual gauge field is on the dual torus and anti-Hermite.
Hence the gauge group is actually $U(k)$.
According to the family index theorem (\ref{f_index}) :
\begin{eqnarray*}
\rank(\Eh)=C_1(E),~~~\Ch_1=\rank(E).
\end{eqnarray*}
This means that
the two-dimensional Nahm transformation
exchanges the rank of gauge group and the first Chern number,
which is consistent with T-duality transformation of D0-D2
brane system.

\vspace{2mm}

In the general arguments above, we assumed the existence of vector bundles 
with $\ker\cD_\xi=0$. Such a vector bundle actually exists for each 
topological charge with positive first Chern number. 
It can be described by a simple solution which describes $k$ D0-branes 
as flux on $N$ D2-branes:
\begin{eqnarray}
&&\Omega_{1}=e^{ikx_2/N}U\ ,~~~ \Omega_{2}=V\ ,~~~ 
 UV=e^{-2\pi ik/N}VU,\label{pm2dim}\\
&&A_1=0,~~~A_2=-\frac{i}{2\pi}\frac{k}{N}x_1,
~~~F_{12}=-\fr{i}{2\pi}\fr{k}{N},~~~C_1=\fr{i}{2\pi}\int_{T^2}d^2x~F_{12}=k.
\label{gf2dim}
\end{eqnarray}
Here the matrices $U$ and $V$ are $N\times N$ matrices
defined by
\begin{eqnarray}
\label{uv}
U_{uv}=\delta_{uv} e^{\frac{2\pi i k u}{N}},~~~V_{uv}
=\delta_{u+1,v}
+\delta_{uN}\delta_{v1}. 
\end{eqnarray}
The patching matrices (transition functions)
$\Omega_{1}(x_1,x_2),~\Omega_{2}(x_1,x_2)$ 
specify the topology of the bundle
and act on the section $\psi$ 
in the fundamental representation of the gauge group as 
\begin{eqnarray}
\label{compa}
\Omega_{1}(x_1,x_2)\psi(x_1,x_2)=\psi(x_1+2\pi,x_2),~~~
\Omega_{2}(x_1,x_2)\psi(x_1,x_2)=\psi(x_1,x_2+2\pi).
\end{eqnarray}
In fact, the patching matrices (\ref{pm2dim}) are defined 
so that the cocycle condition for the vector bundle holds 
\begin{equation*}
\Omega_{1}^{-1}(x_1+2\pi,x_2)\Omega^{-1}_{2}(x_1+2\pi,x_2+2\pi)
\Omega_{1}(x_1,x_2+2\pi)\Omega_{2}(x_1,x_2)=1, 
\end{equation*}
and the covariant derivatives (\ref{gf2dim}) also satisfy the 
following compatibility conditions: 
\begin{eqnarray*}
&&(\partial_\mu+A_\mu)(x_1+2\pi,x_2)
=\Omega_1(x_1,x_2)(\partial_\mu+A_\mu)(x_1,x_2)\Omega^{-1}_1(x_1,x_2),\\
&&(\partial_\mu+A_\mu)(x_1,x_2+2\pi)
=\Omega_2(x_1,x_2)(\partial_\mu+A_\mu)(x_1,x_2)\Omega^{-1}_2(x_1,x_2)
\end{eqnarray*}
for $\mu=1, 2$. 
The general form of the section is given by \cite{GRT, Ho, MoZu}
\begin{eqnarray}
\label{general}
\psi_u(x_1,x_2)=\sum_{s\in \Z}\sum_{p=1}^{k}
\exp\left[ix_1 \left\{
\fr{k}{N}\left(\fr{x_2}{2\pi}+u+Ns\right)+p\right\}\right]
\phi^p\left(\frac{x_2}{2\pi}+u+Ns+\fr{N}{k}p\right).
\end{eqnarray}
One can confirm that this actually satisfies Eq. (\ref{compa}) 
for each $u$.

This section actually has no $\ker\cD_\xi$. 
Now let us solve the Dirac equation (\ref{dirac_02}) 
and give the form of the dual bundle. 
The zero mode have to have the general form (\ref{general})
and satisfies the Dirac equation and the normalized condition.
The solution is 
\begin{eqnarray}
\label{zero_torus}
\psi_u^p(\xi,x)
&=&\left(\fr{N}{2\pi k}\right)^{\qua}
\sum_{s\in \Z} \exp\left[ix_1
\left\{\frac{k}{N}\left(\frac{x_2}{2\pi}+u+Ns\right)+p
\right\}\right]\nn
&&~~~~~~~~~~~~~~\times 
\exp\left[-2\pi i\xi_2\left\{\fr{x_2}{2\pi}+u+Ns-\frac{N}{k}(\xi_1-p)
\right\}\right]\nn
&&~~~~~~~~~~~~~~\times 
\exp\left[-\pi\frac{k}{N}\left\{\frac{x_2}{2\pi}+u+Ns
-\fr{N}{k}(\xi_1-p)\right\}^2\right].
\end{eqnarray}
This solution have several interesting points.
First, the summation over $p$ in the general section (\ref{general})
is dropped out in the zero-mode because of the normalization condition
and hence label $p$ runs just from 1 to $k$, 
which suggests the index theorem on the number of Dirac
zero mode. Second, if we take the opposite sign for the first Chern
number, 
then the Gaussian factor in the third line of Eq. (\ref{zero_torus})
diverges and there is no normalized zero-mode of Dirac operator $\cDb$.
Instead in this case, there is $k$ normalized zero mode of $\cD$,
which also clearly suggests the index theorem.

Finally we get the dual gauge field from Eq. (\ref{gauge_02}): 
\begin{eqnarray*} 
\hat{A}_1=2\pi i\frac{N}{k}\xi_2,~~~\hat{A}_2=0,
~~~\Fh_{12}=-2\pi i\frac{N}{k},
~~~\Ch_1=\fr{i}{2\pi}\int_{\Th^2}d^2\xi~\Fh_{12}=N .
\end{eqnarray*}
where the patching matrices are
\begin{eqnarray}
\label{dual_patch}
\Omegah_{1}=\hat{V}\ ,~~~ \Omegah_{2}=e^{-2\pi iN\xi_1/k}\hat{U}.
\end{eqnarray}
where $\hat{U}, \hat{V}$ are $k\times k$ matrices like (\ref{uv}) such as
$\hat{U}\hat{V}=e^{- 2\pi i\frac{N}{k}}\hat{V}\hat{U}$. 
We also note that the result is already expected at the stage of 
the zero mode (\ref{zero_torus}). It is in fact 
%
compatible with the dual patching matrices (\ref{dual_patch}) except for
the factor $e^{ix_2}$ which disappears in the integration over torus $T^2$.

Thus, we can show that the vector bundles defined by Eq. (\ref{pm2dim}) and 
(\ref{gf2dim}) are transformed to those of the same type. 
In particular, 
%
the transformation exchange the rank of gauge group
and the first Chern number beautifully.
As stated above, 
one can also begin with vector bundles with negative first Chern numbers. 
In this case, one obtains some systems on \=D2-branes. 
Only the trivial bundle, 
that is, the vector bundle with zero first Chern number, 
can not be transformed in the context of these differential geometry. 
In fact, from the index theorem the rank of the dual `bundle' is zero. 
This is a coherent sheaf on the dual torus and can be treated in the 
framework of the algebraic geometry.


\vspace{2mm}

There are some simple generalizations.
Let $E_{(N,k)}$ and $\Eh_{(k,N)}$
be the bundle in question 
on torus $T^2$ and on the dual torus $\Th^2$, respectively. 
We can construct 
the product bundle $E_{(N_1,k_1)}\oplus\cdots\oplus E_{(N_n,k_n)}$.
In this case, the curvature is not a scalar matrix in general 
and the corresponding D0-brane is not regarded as a flux.
Especially if $k_i>0,~k_j<0$ for some $i,j$, then
the D-brane configuration contains $k_i$ D0-branes and $k_j$
anti-D0-brane. In fact, 
when one consider the fluctuation around the solution, 
the fluctuation of the fields between the D-branes $E_{(N_i,k_i)}$ and 
$E_{(N_j,k_j)}$ have negative mass square, 
which implies that the fluctuation includes tachyon modes 
and the system is unstable.
\footnote{The fluctuation spectrum is studied in
detail in, for example, \cite{vanBaal2, Troost}. 
Similar situations are discussed in noncommutative cases in \cite{Ko}.} 
It is interesting that we can take the explicit T-duality transformation
of such D\=D systems in the context of the gauge theories on D-branes
in the present way.
(See also \cite{Hori}.)

\section{Gauge fields on $T^4$ and T-duality}

In this section, we discuss anti-self-dual 
gauge fields on four-tori $T^4$ and 
gauge theoretical descriptions of T-duality transformation
of D0-D4 brane systems which is known as Nahm transformation.
General scheme of the transformation is the same as that of D0-D2.
The important point is that this transformation
preserves the self-duality of the gauge fields.

In this case, we have to consider usual four-dimensional theory.
As the previous section, 
the Dirac operator with real parameter is defined as follows:
\begin{eqnarray*}
\Dsl_\xi &:=& \gamma_\mu \otimes D_{\mu}(\xi)=
\left(
\ba{cc}
0&\cD_\xi\\\cDb_\xi&0
\ea
\right)\\
\Dsl_\xi^2
&=&
\left(
\ba{cc}
\cD_\xi\cDb_\xi&0\\0&\cDb_\xi\cD_\xi
\ea
\right)
=
\left(
\ba{cc}
D^2+\eta^{(-)\mn} F_\mn&0\\0&D^2+\eta^{(+)\mn} F_\mn
\ea
\right),
\end{eqnarray*}
where $D_{\mu}(\xi)=\del_\mu+A_\mu-i\xi_\mu$ and the gamma matrices are
\begin{eqnarray*}
\gamma_\mu&:=&\left(\ba{cc}0&e_\mu \\ \bar{e}_\mu &0\ea\right).
\end{eqnarray*}
The Euclidean 4-dimensional Pauli matrices are defined by
\begin{eqnarray*}
e_\mu:=(-i\sigma_i,1),~~~\bar{e}_\mu=(i\sigma_i,1)
\end{eqnarray*}
and satisfy the following relations:
\begin{eqnarray*}
\bar{e}_\mu e_\nu=\delta_\mn +i\eta_\mn^{i(+)}\sigma_i,~~~
e_\mu \bar{e}_\nu=\delta_\mn +i\eta_\mn^{i(-)}\sigma_i.
\end{eqnarray*}
Here $\eta_\mn^{i(\pm)}$ are called 't Hooft symbol and 
are anti-symmetric and (anti-)self-dual.

First let us consider four-dimensional Dirac operator
with real parameter $\xi_\mu$.
Suppose that the gauge group is $U(N)$ and 
the gauge field is anti-self-dual whose instanton number $C_2$
(the second Chern number) is $k$.

Here we suppose that the Dirac operator $\cD_\xi$ has no smooth
solution.
Then index theorem implies that the Dirac operator $\cDb_\xi$ has independent
normalized $k$ solutions:
\begin{eqnarray}
\label{dirac_04}
\cDb_\xi \psi =0,
\end{eqnarray}
where the each column corresponds to the independent $k$ solution
and hence $\psi$ can be considered as $2N\times k$ matrix.
Then we can construct the gauge field
on the dual torus $\Th^4$ just as in two-dimensional case:
\begin{eqnarray*}
\Ah_\mu=\int_{T^4}dx^4~\psi^\dagger \delh_\mu \psi.
\end{eqnarray*}
The second Chern number of the gauge field $\Ah$ is $N$.
This transformation is in fact one-to-one
and called Nahm transformation.

In the following, we construct various (anti-)self-dual gauge field
and take the Nahm transformation by solving the Dirac equations.

\vspace{2mm}

The first example is very simple. It is essentially given by 
the tensor product of the previous example on two-dimensional torus. 

The anti-self-dual gauge field is given as follows:
\begin{eqnarray}
\label{exmp1}
A_1=0,~~~A_2=-\frac{i}{2\pi}\frac{k}{N}x_1\otimes\1_{N\times N},~~~
A_3=0,~~~A_4=\1_{N\times N}\otimes\frac{i}{2\pi}\frac{k}{N}x_3.
\end{eqnarray}
The curvature is
\begin{eqnarray*}
F_{12}=-F_{34}=-\fr{i}{2\pi}\fr{k}{N}\1_{N\times N}\otimes\1_{N\times N}.
\end{eqnarray*}
and the second Chern number is $k^2$.
This bundle is constructed as 
tensor-like product $E_{(N,k)}\otimes E_{(N,-k)}$.
Because of the opposite twist, 
self-duality of the gauge field are realized.
The gauge group is considered as $U(N^2)$.

The Dirac zero mode is essentially the product of two-dimensional case:
\begin{eqnarray}
\label{zero_torus2}
\psi_{uu^\prime}^{pp^\prime}(\xi,x)
&=&\left(\frac{N}{2\pi k}\right)^{\half}
\sum_{s,t\in \Z} e^{ix_1(\frac{k}{N}(\frac{x_2}{2\pi}+u+Ns)+p)}
e^{-2\pi i\xi_2 (\fr{x_2}{2\pi}+u+Ns-\frac{N}{k}(\xi_1-p))}
e^{-\frac{\pi k}{N}(\frac{x_2}{2\pi}+u+Ns-\fr{N}{k}(\xi_1-p))^2}\nonumber\\
&&~~~\times e^{-ix_3(\frac{k}{N}(\frac{x_4}{2\pi}+u^\prime+Nt)+p^\prime)}
e^{2\pi i\xi_4
(\fr{x_4}{2\pi}+u^\prime+Nt-\frac{N}{k}(\xi_3-p^\prime))}
e^{-\frac{\pi k}{N}(\frac{x_4}{2\pi}+u^\prime+Nt-\fr{N}{k}(\xi_3-p^\prime))^2}.
\end{eqnarray}
Then the dual gauge field becomes
\begin{eqnarray*}
\hat{A}_1=2\pi i\frac{N}{k}\xi_2\otimes\1_{k\times k},~~~\hat{A}_2=0,~~~
\hat{A}_3=\1_{k\times k}\otimes-2\pi i\frac{N}{k}\xi_4,~~~\hat{A}_4=0.
\end{eqnarray*}
whose second Chern number is $N^2$ and gauge group is $U(k^2)$.
So D0-brane charge and D4-brane charge are exchanged.  

\vspace{2mm}

As discussed previously, there is a simple generalization of this:
the Nahm transformation 
of $(E_{(N_1,k_1)}\otimes E_{(N_1,-k_1)})\oplus \cdots \oplus 
(E_{(N_n,k_n)}\otimes E_{(N_n,-k_n)})$
where the rank of the gauge group is $\sum_{i=1}^{n}n^2_i$
and the second Chern number is $\sum_{i=1}^{n}k^2_i$.
If we take $\sum_{i=1}^n k_i=0$, then 
the first Chern number becomes all zero and
D2-brane charges disappear. 
The discussion in the  $n=2, N_1=1, N_2=1, k_1=1, k_2=-1$ case
coincides with that in \cite{vanBaal3}.
Alternatively, if we consider a self dual configuration as 
$E_{(N_1,k_1)}\otimes E_{(N_1,k_1)}$, 
it is transformed to the systems on \=D4-brane. 
Moreover, if one consider the direct sum of a 
anti-self-dual configuration and a self dual configuration, 
one can see that the fluctuations corresponding to the modes between them 
are negative and include the tachyonic modes 
similarly as discussed on the two-dimensional tori.

\vspace{2mm}

The previous example has the separate form between
1-2 direction and 3-4 direction.
But we can treat the mixed version as follows.
For simplicity, we focus on $G=U(1), C_2=2$.

The gauge field is given by
\begin{eqnarray*}
A_1=0,~~~A_2=-\frac{i}{2\pi}
(x_1-x_3),~~~A_3=0,~~~
A_4=\frac{i}{2\pi}
(x_1+x_3).
\end{eqnarray*}
The field strength is 
\begin{eqnarray*}
F_{12}=-F_{34}=\fr{i}{2\pi}
, ~~~F_{14}=-F_{23}=\fr{i}{2\pi}.
\end{eqnarray*}
Now $F_{14}$ part contributes more to the second Chern number.
Hence the number becomes two.

In this case, the corresponding section is of the form
\begin{eqnarray}
 \psi(x_1,x_2,x_3,x_4)
 &=&\sum_{s,t\in\Z, p\in\Z_2}
 \exp{\left[i(x_1-x_3)\left(\frac{x_2}{2\pi}+s+\frac{p}{2}\right)
      -i(x_1+x_3)\left(\frac{x_4}{2\pi}+t+\frac{p}{2}\right)\right]}\nn
 &&~~~~~~~~~~\times\phi^p\left(\frac{x_2}{2\pi}+s+\frac{p}{2},
 \frac{x_4}{2\pi}+t+\frac{p}{2}\right)\ .
 \label{ms1}
\end{eqnarray}
This can also be written as 
\begin{eqnarray*}
 \psi(x_1,x_2,x_3,x_4)
 &=&\sum_{s,t\in\Z, p\in\Z_2}
 \exp{\left[i x_1\left(\frac{x_2-x_4}{2\pi}+s'\right)
      -i x_3\left(\frac{x_2+x_4+p}{2\pi}+t'\right)\right]}\nn 
 &&~~~~~~~~\times
\phi^p\left(\frac{x_2}{2\pi}+s+\frac{p}{2},\frac{x_4}{2\pi}
+t+\frac{p}{2}\right)\ .
\end{eqnarray*}
where $s':=s-t$ and $t':=s+t+p$ run over $\Z^2$ densely. 
The fact explains the reason 
why $p$ is necessary in Eq. (\ref{ms1}). 
%
In the process of solving the Dirac equation $\cDb_\xi\psi=0$, 
one can see that the solutions are of the form
\begin{eqnarray*}
\psi^p= \bp \psi^p_+\\ \psi^p_- \ep
 &=& \bp 1 \\ c \ep
 \sum_{s,t\in\Z}
 \exp{\Big[i(x_1-x_3)
\frac{x'_2}{2\pi}
      -i(x_1+x_3)
\frac{x'_4}{2\pi}
\Big]}\nn
 &&~~~~~\times
 \exp{\Big[-(a_2(x'_2)^2+a_4(x'_4)^2+a_{24}x'_2 x'_4
+b_2 x'_2+ b_4 x'_4)\Big]} 
\end{eqnarray*}
where $x_2':=x_2+2\pi(s+p/2)$, 
$x_4':=x_4+2\pi(t+p/2)$, 
$c, a_2, a_4, a_{24}, b_2, b_4\in\C$ and especially 
$\Re(a_2)>0$, $\Re(a_4)>0$. 
Solving the differential equations leads quadratic equations. 
Especially, the quadratic equation for $c$ is obtained:
\begin{equation*}
 c^2+2c-1=0.
\end{equation*}
This equation has the two solutions: $c=-1\pm\sqrt{2}$.
On the other hand, 
both $a_2$ and $a_4$ are positive only when $c=-1+\sqrt{2}$
because of $a_2+a_4=(1+c)/2\pi$.
For fixed $c$, the rest variables $a_2, a_4, a_{24}, b_2, b_4$ 
are determined uniquely. 
Thus we can confirm that there are two orthogonal Dirac zero-modes 
$(\psi^p_+,\psi^p_-)$ for $p=0, 1$. 

Next, the $\xi$ dependence is determined by 
normalizing these zero-modes.
We get
\begin{eqnarray*}
&& \bp \psi^p_+\\ \psi^p_- \ep
 \propto \bp 1 \\ c \ep
 \sum_{s,t\in\Z}
 \exp{\Big[i(x_1-x_3)
\frac{x'_2}{2\pi}
      -i(x_1+x_3)
\frac{x'_4}{2\pi}
\Big]}\nn
&&  ~~~~~~~\times
  \exp{\Big[-\frac{1}{2} \Re(X^t)\bp 2a_2 & a_{24}\\ a_{24} & 2a_4\ep \Re(X)
  -i\Re(X^t)\bp 2a_2 & a_{24}\\ a_{24} & 2a_4\ep \Im(X)\Big]}
\end{eqnarray*}
where 
\begin{equation*}
 X=\bp x'_2
\\x'_4
\ep
  +2\pi
 \bp 1+c & 1-c\\ 1-c & 1+c \ep^{-1}
 \bp -(\xi_1+i\xi_2)+c(\xi_3+i\xi_4)\\ 
     (\xi_3-i\xi_4)+c(\xi_1-i\xi_2)\ep\ .
\end{equation*}
Then the dual gauge field is obtained as 
\begin{equation*}
 A_1=\frac{2\pi i}{2}(-\xi_2+\xi_4)\1_{2\times 2},~~~A_2=0,~~~
A_3=-\frac{2\pi i}{2}(\xi_2+\xi_4)\1_{2\times 2},~~~A_4=0\ .
\end{equation*}
One can then confirm that the topological numbers agree with the 
ones expected from the T-duality or the family index theorem (\ref{f_index}), 
in particular the rank is two and $\Ch_2=1$.


\section{Generalizations}

We can easily generalize the previous discussions to
higher-dimensional case.
The family index theorem (\ref{f_index}) holds in every even dimensions
and the zero-mode bundle can be constructed in fact even if we 
suppose that there is no kernel of $\cD$.
Hence we can perform the explicit
higher-dimensional Nahm transformation for the constant
curvature gauge field in \cite{Taylor, Troost}.
The zero mode can be constructed as the product of the
two-dimensional zero-mode (\ref{zero_torus}) 
as in four dimensional one (\ref{zero_torus2}).

Note that we can construct any other constant curvature bundles 
in terms of Heisenberg modules \cite{KoSc} 
and take their Nahm transformations for arbitrary dimensional tori. 
Heisenberg modules are known as projective modules 
(noncommutative analogue of vector bundles) over noncommutative tori
with constant curvature connections
\cite{KoSc}, but they can also be used in commutative cases. 
Recall the constant curvature bundles we have used in the body 
of this paper. 
The forms of their sections, 
for example Eq. (\ref{general}) and Eq. (\ref{ms1}), are 
essentially determined by $\phi^p$ for fixed Chern characters. 
The $\phi^p$'s are nothing but the Heisenberg modules. 
Generally,  
the Heisenberg modules over $n$ dimensional (noncommutative) tori 
are described as functions on $\R^d\times \Z^{d'}\times F$ where 
$F$ is a finite group and $2d+d'=n$. 
They are specified by defining the action of the generators of 
$C^\infty(T^n)$. 
Here $\Z^{d'}$ part corresponds to that the bundle is trivial 
for the corresponding $d'$ directions of the torus. 
This means that its Nahm dual module is 
not described by a vector bundle. 
Hence we consider the case 
$\R^d\times F$ where $n=2d$. 
Then one can see that the $\phi^p$ in Eq. (\ref{general}) is the case 
$d=1, F=\Z_k$. 
Similarly, the example (\ref{exmp1}) corresponds to the case 
$d=2, F=\Z_k\times\Z_k$. 
$\phi^p$ in Eq. (\ref{ms1}) and its dual bundle are then the case 
$d=2, F=1$ and $d=2, F=\Z_2$, respectively. 
Any Heisenberg module has a constant curvature. 
Moreover, when one construct the twisted bundles corresponding to the 
Heisenberg module as in our examples, 
the constant curvature on the Heisenberg module is compatible with 
that on the twisted bundle. 
Namely, our examples of the Nahm transformations are essentially 
those accomplished by using Heisenberg modules and 
the extensions to any other constant curvature bundles can also be 
done in terms of Heisenberg modules.

\section{Conclusions and Discussions}

We studied the transformations of gauge fields which 
correspond to the T-duality transformations 
on even dimensional tori. 
On two-dimensional tori, after expanding general arguments of 
the transformations, 
we presented the explicit example, where 
we considered a constant curvature bundle for any topological 
number and took the Nahm-like transformation. 
The dual bundle was confirmed to be the same type as the original 
constant curvature bundle, and the result certainly agrees with the 
T-duality transformations. 
On four-dimensional tori, we considered the instanton configurations 
of gauge fields and their transformations, which are just the 
Nahm transformations. 
Applying the results to the two-dimensional situations, 
we constructed various constant curvature bundle and took their 
Nahm transformations explicitly. 
One can also consider the direct sum of these constant curvature bundle 
and their Nahm-like transformations. 
We saw that these situations generally express the 
T-duality transformations of D\=D systems. 
Finally we commented about higher dimensional extensions. 
In particular, for any topological number the corresponding 
constant curvature bundle 
and their Nahm-like transformations can be obtained in terms of 
Heisenberg modules.

One of the future direction is noncommutative extensions of 
the Nahm transformation. 
A noncommutative Nahm transformation is discussed formally 
in \cite{ANS}, however the noncommutative Nahm transformation have not 
ever performed. 
{}From the viewpoint along this paper, 
one may deal with the noncommutative version of the twisted bundles
discussed in \cite{Ho, MoZu}. 
However, it is difficult to solve the Dirac equation consistently 
since it is a differential equation on noncommutative space. 
One may also consider Heisenberg modules. 
However, one arrives at the same problem if 
one tries to obtain the Dirac zero modes by 
solving a differential equation. 
Another approach is to define the tensor product between 
two Heisenberg modules directly like as in \cite{DiSc, Kajiura2}. 
Also, noncommutative versions of index theorems may be useful. 
Anyway, this problem seems to be relevant to the 
Morita equivalence \cite{KoSc} of D\=D system on noncommutative tori
\cite{BKMT,KMT} from physical viewpoints. 
We would like to report such directions elsewhere.

\vspace{2mm}
\begin{center}
\noindent{\large \textbf{Acknowledgments}}
\end{center}

We would greatly like to thank T.~Takayanagi
for collaboration at the early stage of this work.
We are very grateful to A.~Kato and Y.~Matsuo 
for advice,
and P.~van Baal, M.~Furuta, A.~Schwarz and Y.~Terashima
for useful comments. 
We also acknowledge the Summer Institute
2001 at Fuji-Yoshida where this work began,
and the YITP at Kyoto university during the YITP workshop YITP-W-02-04
where this work was completed.
The work of M.H. was supported in part by
the Japan Securities Scholarship Foundation (\#12-3-0403).
The work of H.K. was supported 
by JSPS Research Fellowships for Young
Scientists.

\appendix


\begin{thebibliography}{99}

\bibitem{tHooft}
G.~'t Hooft,
Nucl.\ Phys.\ B {\bf 153} (1979) 141.

\bibitem{Gonzalez}
A.~Gonzalez-Arroyo,
``Yang-Mills fields on the 4-dimensional torus. (Classical theory),''
hep-th/9807108.

\bibitem{Mukai} 
S.~Mukai,
Nagoya Math. J. {\bf 81} (1981) 153.

\bibitem{Schenk}
H.~Schenk,
Commun.\ Math.\ Phys.\  {\bf 116} (1988) 177.

\bibitem{BvB}
P.~J.~Braam and P.~van Baal,
Commun.\ Math.\ Phys.\  {\bf 122} (1989) 267.

\bibitem{DoKr} 
S.~K.~Donaldson and P.~B.~Kronheimer, 
{\it The Geometry of Four-Manifolds} 
(Oxford UP, 1990) 
[ISBN/0-19-850269-9].

\bibitem{ADHM}
M.~F.~Atiyah, N.~J.~Hitchin, V.~G.~Drinfeld and Y.~I.~Manin,
Phys.\ Lett.\ A {\bf 65} (1978) 185.

\bibitem{Nahm}
W.~Nahm,
Phys.\ Lett.\ B {\bf 90} (1980) 413;
``The construction of all self-dual multimonopoles by the ADHM method,''
{\it Monopoles in Quantum Field Theory}
(1982) 87 [ISBN/9971-950-29-4].

\bibitem{Harvey}
J.~A.~Harvey,
``Komaba lectures on noncommutative solitons and D-branes,''
[hep-th/0102076].

\bibitem{AtSi}
M.~F.~Atiyah and I.~M.~Singer,
Annals Math.\  {\bf 93} (1971) 119;
Annals Math.\  {\bf 93} (1971) 139.

\bibitem{GRT}
O.~J.~Ganor, S.~Ramgoolam and W.~I.~Taylor,
Nucl.\ Phys.\ B {\bf 492} (1997) 191
[hep-th/9611202].

\bibitem{Ho}
P.~M.~Ho,
Phys.\ Lett.\ B {\bf 434} (1998) 41
[hep-th/9803166].

\bibitem{MoZu}
B.~Morariu and B.~Zumino,
hep-th/9807198.

\bibitem{Hori}
K.~Hori,
Adv.\ Theor.\ Math.\ Phys.\  {\bf 3} (1999) 281
[hep-th/9902102].

\bibitem{vanBaal2}
P.~van Baal,
Commun.\ Math.\ Phys.\  {\bf 94} (1984) 397.

\bibitem{Troost}
J.~Troost,
Nucl.\ Phys.\ B {\bf 568} (2000) 180
[hep-th/9909187].

\bibitem{Ko}
A.~Konechny,
JHEP {\bf 0203} (2002) 035
[hep-th/0112189].

\bibitem{vanBaal3}
P.~van Baal,
Nucl.\ Phys.\ Proc.\ Suppl.\  {\bf 49} (1996) 238
[hep-th/9512223].

\bibitem{Taylor}
W.~I.~Taylor,
Nucl.\ Phys.\ B {\bf 508} (1997) 122
[hep-th/9705116].

\bibitem{KoSc}
A.~Konechny and A.~Schwarz,
``Introduction to M(atrix) theory and noncommutative geometry,''
hep-th/0012145.

\bibitem{ANS}
A.~Astashkevich, N.~Nekrasov and A.~Schwarz,
Commun.\ Math.\ Phys.\  {\bf 211} (2000) 167
[hep-th/9810147].

\bibitem{DiSc}
M.~Dieng and A.~Schwarz,
``Differential and complex geometry of two-dimensional noncommutative tori,''
math.QA/0203160.

\bibitem{Kajiura2}
H.~Kajiura,
JHEP {\bf 0208} (2002) 050
[hep-th/0207097].

\bibitem{BKMT}
I.~Bars, H.~Kajiura, Y.~Matsuo and T.~Takayanagi,
Phys.\ Rev.\ D {\bf 63} (2001) 086001
[hep-th/0010101].

\bibitem{KMT}
H.~Kajiura, Y.~Matsuo and T.~Takayanagi,
JHEP {\bf 0106} (2001) 041
[hep-th/0104143].

\end{thebibliography}
\end{document}